\documentstyle[12pt]{article}
\title{Predicting the neutrino spectrum
in minimal SO(10) Grand Unification}
\author{L.\ Lavoura\\
\small Department of Physics, Carnegie-Mellon University, \\
\small Pittsburgh, Pennsylvania 15213, U.S.A.}
\begin{document}
\maketitle
\begin{abstract}
I analyse the model of Babu and Mohapatra
for the fermion mass matrices in minimal SO(10).
Those authors have not considered
the whole variety of fits to the experimental data
possible with their mass matrices.
Consequently,
their predictions for the neutrino spectrum are incomplete.
I survey various types of neutrino spectra possible in their model.
\end{abstract}

\vspace{5mm}

Babu and Mohapatra (BM) \cite{bm} have recently noted that
in a minimal SO(10) model the lepton mass matrices
can be written as functions of the quark mass matrices:
\begin{eqnarray}
M_l & = & a M_u + b M_d\, ,
\nonumber\\
M_{\nu}^D & = & (- b - 2) M_u + \frac{(b + 3)(1 - b)}{a} M_d\, ,
\label{eq:matrices}\\
M_{\nu}^M & \propto & M_u + \frac{b - 1}{a} M_d\, .
\nonumber
\end{eqnarray}
$ M_u $ and $ M_d $ are the mass matrices of the up-type and down-type quarks,
respectively.
$ M_l $ is the mass matrix of the charged leptons.
$ M_{\nu}^D $ and $ M_{\nu}^M $ are the Dirac and Majorana mass matrices,
respectively,
of the neutrinos.
The effective mass matrix of the light neutrinos is,
in the seesaw \cite{seesaw} approximation,
\begin{equation}
M_{\nu}^{{\rm eff}} = - M_{\nu}^D M_{\nu}^M (M_{\nu}^D)^T\, .
\label{eq:effective}
\end{equation}
All the mass matrices,
in Eqs.~\ref{eq:matrices} as well as in Eq.~\ref{eq:effective},
are real and symmetric.
$ a $ and $ b $ are real numbers.\footnote{The scheme
proposed by BM is more general,
in that it also includes two phases,
which render the mass matrices complex.
But BM have only analysed in detail the case of real mass matrices
and I shall also consider only that restriction of their scheme.}

The model of BM is attractive because of its minimality in
the content of scalar representations.
BM only use one {\bf 10} and one {\bf 126} of Higgs scalars
coupling to the fermions.
These representations are necessary anyway:
the {\bf 126} is needed to give Majorana masses to the neutrinos
and to generate the seesaw mechanism;
the {\bf 10} is needed in order to obtain $ m_b \approx m_{\tau} $
at the breaking scale of the Pati--Salam \cite{pati}
subgroup SU(4)$_{PS}$ of SO(10).
The predictive power of the BM model originates
in the small number of Yukawa-coupling matrices,
not in {\it ad hoc} assumptions,
like discrete symmetries leading to vanishing Yukawa couplings,
which are frequent in other models \cite{models}.
BM have noted that the {\bf 126},
besides giving Majorana masses to the right-handed neutrinos,
also introduces welcome corrections to the other mass matrices.
Therefore the BM model is worth especial attention.
My purpose in this Brief Report is to fit the known data
on the quark masses and mixings,
and on the charged-lepton masses,
by mass matrices of the BM type,
and then to find out which kinds of neutrino spectra
are possible in the model.

The BM model leads to two conditions on the quark masses and mixings,
and on the charged-lepton masses.
The first condition is that,
as the quark mass matrices $ M_u $ and $ M_d $ are real,
the quark mixing matrix $ V $ is real.
It is well known \cite{branco} that,
once $ |V_{us}| $,
$ |V_{cb}| $,
$ |V_{ub}| $ and $ |V_{td}| $ are given,
the quark mixing matrix is completely determined.
I have used the following fixed values in my fits:
$ |V_{us}| = 0.22 $,
$ |V_{cb}| = 0.046 $,
and $ |V_{ub}| = 0.004 $.
The condition that $ V $ is orthogonal
then leads to two possible values for $ |V_{td}| $:
either $ |V_{td}| = V_{{\rm min}} = 0.006223 $
or $ |V_{td}| = V_{{\rm max}} = 0.014017 $.
(For $ V_{{\rm min}} < |V_{td}| < V_{{\rm max}} $ the matrix $ V $
is unitary but not real;
for $ |V_{td}| < V_{{\rm min}} $ or $ |V_{td}| > V_{{\rm max}} $
the matrix $ V $ is not unitary.)
If one accepts the standard-model explanation
of $B_d$-$\overline{B_d}$ mixing
by means of the box diagram \cite{bd}
and uses a top-quark mass of 150 GeV (see below),
then $ |V_{td}| = V_{{\rm max}} $ corresponds to
$ \tilde{f}_B \equiv f_B \sqrt{B_B \eta_B} = 116 $ MeV,
a quite reasonable value.
$ |V_{td}| = V_{{\rm min}} $ corresponds to $ \tilde{f}_B = 262 $ MeV,
a rather high value,
but still acceptable.

The second condition can be derived
from the first of Eqs.~\ref{eq:matrices}.
Obviously,
\begin{eqnarray}
{\rm tr} M_l & = & a\, {\rm tr} M_u + b\, {\rm tr} M_d\, ,
\nonumber\\
{\rm tr} M_l^2 & = & a^2\, {\rm tr} M_u^2 + 2 a b\, {\rm tr} (M_u M_d)
+ b^2\, {\rm tr} M_d^2\, ,
\label{eq:traces}\\
{\rm tr} M_l^3 & = & a^3\, {\rm tr} M_u^3
+ 3 a^2 b\, {\rm tr} (M_u^2 M_d)
+ 3 a b^2\, {\rm tr} (M_u M_d^2) + b^3\, {\rm tr} M_d^3\, .
\nonumber
\end{eqnarray}
Because $ M_u $ and $ M_d $ are Hermitian,
$ {\rm tr} (M_u^p M_d^q) = \sum_{i = u, c, t}\,
\sum_{j = d, s, b} m_i^p m_j^q |V_{i j}|^2 $.
Therefore,
all the traces in the right-hand sides of Eqs.~\ref{eq:traces}
are observables.
The traces in their left-hand sides are functions
of the charged-lepton masses.
We thus have three equations
for the two unknowns $ a $ and $ b $.
{}From the first equation,
$ b $ is determined as a function of $ a $.
Substituting the result in the second and third equations,
we obtain a set of a quadratic equation and a cubic equation for $ a $:
\begin{eqnarray}
x_1 a^2 + 2 x_2 a + x_3 & = & 0\, ,
\nonumber\\
y_1 a^3 + 3 y_2 a^2 + 3 y_3 a + y_4 & = & 0\, ,
\label{eq:aequations}
\end{eqnarray}
where the coefficients $ x_{1,2,3} $
and $ y_{1,2,3,4} $ are functions of traces,
and are therefore observables.
Eqs.~\ref{eq:aequations} fix $ a $ to be
\begin{equation}
a = \frac{3 x_1 x_3 y_2 - 2 x_2 x_3 y_1 - x_1^2 y_4}{4 x_2^2 y_1
+ 3 x_1^2 y_3 - 6 x_1 x_2 y_2 - x_1 x_3 y_1}\, .
\label{eq:asolution}
\end{equation}
Besides,
because one has two equations for only one unknown,
the coefficients of Eqs.~\ref{eq:aequations}
must satisfy the following condition:
\begin{eqnarray}
x_1^3 y_4^2 + 9 x_1^2 x_3 y_3^2 - 6 x_1^2 x_2 y_3 y_4
&   &
\nonumber\\
-6 x_1^2 x_3 y_2 y_4 + 6 x_1 x_2 x_3 y_1 y_4 - 18 x_1 x_2 x_3 y_2 y_3
&   &
\nonumber\\
+ 9 x_1 x_3^2 y_2^2 - 6 x_1 x_3^2 y_1 y_3 + 12 x_1 x_2^2 y_2 y_4
&   &
\nonumber\\
- 6 x_2 x_3^2 y_1 y_2 - 8 x_2^3 y_1 y_4 + x_3^3 y_1^2
+ 12 x_2^2 x_3 y_1 y_3 & = & 0\, .
\label{eq:condition}
\end{eqnarray}
This is the second condition on the quark masses and mixings,
and on the charged-lepton masses,
to which I referred above.

I have used fixed and exact values for the following masses:
$ m_t = 150 \times 0.506 $ GeV,
$ m_b = 4.25 \times 0.327 $ GeV,
$ m_c = \pm 1.27 \times 0.286 $ GeV,
$ m_u = \pm 5.1 \times 0.273 $ MeV,
$ m_{\tau} = 1.784 \times 0.960 $ GeV,
$ m_{\mu} = \pm 105.6584 \times 0.960 $ MeV
and $ m_e = \pm 510.999 \times 0.960 $ keV,
in which I have used the renormalization factors given by BM.
The masses of the top and bottom quarks may,
without loss of generality,
be set positive.
The masses of all other quarks and of the charged leptons
may have either sign.
However,
I am interested in a picture of the mass matrices
in which the contributions from the {\bf 10} of scalars
dominate over the contributions from the {\bf 126}.
For every fit of the data,
one may make the transformation $ a \rightarrow - a $
and $ b \rightarrow - b $ to obtain another fit,
which has symmetric charged-lepton masses,
but a different neutrino spectrum and a different leptonic mixing matrix.
For each particular fit,
one may separate the contributions of the
{\bf 10} and of the {\bf 126} to $ M_d $ and $ M_l $:
the contribution of the {\bf 10} is $ (3 M_d + M_l) / 4 $,
and the contribution of the {\bf 126} is $ (M_d - M_l) / 4 $.
I have computed the ratio of the highest eigenvalue
of the contribution of the {\bf 10}
to the highest eigenvalue of the contribution of the {\bf 126}.
I have found that,
if $ m_b $ and $ m_{\tau} $ have opposite signs
that ratio is close to one,
and I have discarded those fits.
But,
if $ m_b $ and $ m_{\tau} $ have the same sign,
the ratio is typically somewhere between 5 and 15;
those are the fits that satisfy my (and BM's) prejudices.
Thus,
from each pair of fits related to each other by the symmetrization
$ M_l \rightarrow - M_l $ referred to above,
I only consider the fit in which $ m_{\tau} $ is positive.

My fitting method has been the following.
Having fixed all the quark masses,
together with their signs,
except $ m_s $ and $ m_d $,
and having I looked for pairs of values of $ m_s $ and $ m_d $
such that the condition of Eq.~\ref{eq:condition} is satisfied.
For each such pair of values,
one automatically obtains a fit of the experimental data
by the BM mass matrices,
{\it i.e.},
values of $ a $ and $ b $.
I was careful to restrict my search of $ m_s $ and $ m_d $
to the one-standard-deviation ranges suggested by Gasser and Leutwyler
\cite{gasser}:
$ 18.0 \le |m_s / m_d| \le 21.2 $
and $ 1.63 \le |m_d / m_u| \le 1.89 $
at the renormalization scale 1 GeV.
(Remember that $ |m_u (1\, {\rm GeV})| = 5.1 $ MeV is kept fixed.)
I also used the renormalization factors for $ m_s $ and $ m_d $
suggested by BM.

I found four different types of fits.
All four types of fits have a positive muon mass
and a positive ratio $ m_s / m_d $.
Fits of type 1 have positive charm mass,
all others have $ m_c < 0 $.
Fits of type 2 have positive strange mass,
all others have $ m_s < 0 $.
Fits of type 4 have $ |V_{td}| = V_{{\rm max}} $,
all others have $ |V_{td}| = V_{{\rm min}} $.
For all four types of fits,
the signs of the up mass and of the electron mass
are not very relevant,
solutions being found for any choice of those signs,
and the resulting neutrino spectra being very similar.
For each type of fit and for each choice of signs for $ m_e $ and $ m_u $,
a curve may be drawn in the $ m_s\, {\rm vs.}\, m_d $ plane,
displaying the pairs of values for which the fits are found
(which yield solutions of Eq.~\ref{eq:condition}).
The fits presented by BM in their original work are all of type 4.

I emphasize that I have been very exacting in my assumptions.
I have not allowed any quark masses,
except for $ m_s $ and $ m_d $,
to vary,
and I have used the central values suggested by Gasser and Leutwyler;
I have not allowed the renormalization factors
to vary from the values suggested by BM;
and I have considered only the one-standard-deviation ranges for
$ |m_s / m_d| $ and $ |m_d / m_u| $.
I have checked that if one allows instead those two ratios
to be anywhere in their three-standard-deviation ranges,
one obtains other types of fits to the data,
beyond types 1 to 4,
but I do not consider those other types of fits in this Brief Report.
I also fixed $ V $ except for the ambiguity
$ |V_{td}| = V_{{\rm max}} $ or $ |V_{td}| = V_{{\rm min}} $,
and I have discarded the fits with $ (m_b m_{\tau}) < 0 $.

In the fits of type 4
the variable $ r_2 $ introduced by BM may be infinite
(which corresponds in my notation to $ b = -3 $),
indicating a null contribution of the {\bf 126}
to $ M_u $ and to $ M_{\nu}^D $.
Such a possibility does not arise in the fits of other types.

I now turn to the neutrino spectra.
Having found $ a $ and $ b $ one may compute $ M_{\nu}^{{\rm eff}} $.
Its eigenvalues $ m_3 $,
$ m_2 $ and $ m_1 $,
ordered as $ |m_3| \ge |m_2| \ge |m_1| $,
are proportional to the light-neutrino masses.
{}From the diagonalization of $ M_l $ and of $ M_{\nu}^{{\rm eff}} $
one computes the lepton mixing matrix $ K $.
I find that the four kinds of fits yield four different
types of neutrino spectra.

Fits of types 1 and 2 yield well-defined neutrino spectra,
with small variations of the various parameters
as one lets $ m_s $ and $ m_d $ evolve along the curves
giving the solutions to Eq.~\ref{eq:condition}.
$ K $ is dominated by a small first-second generation mixing
in both cases,
and the ratios of neutrino masses are small.
For fits of type 1 we have
\begin{eqnarray}
23 < |m_3 / m_2| < 27 & , &
\nonumber\\
1.6 < |m_2 / m_1| < 2.1 & , &
\nonumber\\
0.325 < |K_{\nu_1 \mu}| < 0.385 & , &
\label{eq:fit1}\\
0.045 < |K_{\nu_1 \tau}| < 0.054 & , &
\nonumber\\
0.031 < |K_{\nu_2 \tau}| < 0.037 & . &
\nonumber
\end{eqnarray}
For fits of type 2 we have
\begin{eqnarray}
17.9 < |m_3 / m_2| < 18.8 & , &
\nonumber\\
11.2 < |m_2 / m_1| < 12.7 & , &
\nonumber\\
0.091 < |K_{\nu_1 \mu}| < 0.104 & , &
\label{eq:fit2}\\
0.010 < |K_{\nu_1 \tau}| < 0.012 & , &
\nonumber\\
0.014 < |K_{\nu_2 \tau}| < 0.016 & . &
\nonumber
\end{eqnarray}

In contrast with these,
fits of types 3 and 4 lead to less well-defined neutrino spectra.
Fits of type 3 are characterized by very large,
even maximal,
lepton mixing.
They yield
\begin{eqnarray}
1 \le |m_3 / m_2| < 12 & , &
\nonumber\\
24 < |m_3 / m_1| < 33 & , &
\nonumber\\
0.35 < |K_{\nu_1 \mu}| < 0.75 & , &
\label{eq:fit3}\\
0.09 < |K_{\nu_1 \tau}| < 0.13 & , &
\nonumber\\
0.24 < |K_{\nu_2 \tau}| < 0.71 & . &
\nonumber
\end{eqnarray}
When one lets $ m_s $ and $ m_d $ evolve
along the curves giving the solutions of Eq.~\ref{eq:condition},
the mixing of the first two generations,
given by $ |K_{\nu_1 \mu}| $,
diminishes,
while simultaneously the mixing of the last two generations,
given by $ |K_{\nu_2 \tau}| $,
increases.
Also,
$ |m_2 / m_1| $ increases while $ |m_3 / m_2| $ decreases,
and this in such a way that their product remains approximately constant
(see the second of Eqs.~\ref{eq:fit3}).
At a certain point the masses of the two heaviest neutrinos
become equal,
at which point the second and third rows of $ K $
must be interchanged.

Fits of type 4 were the only ones considered by BM.
They are characterized by large ratios among the neutrino masses.
One has
\begin{eqnarray}
700 < |m_3 / m_2| < 2600 & , &
\nonumber\\
20 < |m_2 / m_1| < 30000 & , &
\nonumber\\
0 \le |K_{\nu_1 \mu}| < 0.4 & , &
\label{eq:fit4}\\
0.006 < |K_{\nu_1 \tau}| < 0.060 & , &
\nonumber\\
0.103 < |K_{\nu_2 \tau}| < 0.111 & . &
\nonumber
\end{eqnarray}

Let us confront the various neutrino spectra with the MSW \cite{msw}
explanation of the solar-neutrino deficit.
Starting by the solutions of type 4 (see Eq.~\ref{eq:fit4}),
we notice that the large second-third generation mixing
allows us to make use of the experimental data on
$\nu_{\mu}$--$\nu_{\tau}$ mixing \cite{mutau,future}
to obtain $ |m_3| < 2 $ eV.
Then,
$ |m_3 / m_2| $ must be smaller than 1300
if $ |m_2| $ is not to be too small
for the MSW effect to explain the data.
This is possible for solutions of type 4,
and therefore it is possible to accomodate a $\nu_e$--$\nu_{\mu}$
MSW effect in some of these cases \cite{bm}.
But,
if this occurs,
future experiments (P860) should be able to detect
$\nu_{\mu}$--$\nu_{\tau}$ oscillations,
because $ m_{\nu_{\tau}} $ must be very close to 2 eV.
Another viable possibility
is $\nu_e$--$\nu_{\tau}$ MSW depletion of the solar-neutrino flux,
in which case however all the neutrino masses would be very small,
and uninteresting from the cosmological point of view.
If MSW $\nu_e$--$\nu_{\tau}$ oscillations
explain the solar-neutrino depletion in this type of solutions,
then the $\nu_{\mu}$ mass will be in the range $10^{-12}$--$10^{-11}$ eV,
and this would,
by means of normal vacuum oscillations,
provoke a further depletion of the neutrinos observed
in the gallium experiments \cite{vacuum}.

The MSW effect cannot occur in solutions of type 2
(see Eq.~\ref{eq:fit2}),
because the $\nu_1$--$\nu_3$ mixing is too small
while the $\nu_1$--$\nu_2$ mixing falls in the range between
the small-mixing-angle and the large-mixing-angle fits
of the MSW effect to the data.
Solution 2 is interesting because of its very small
$\nu_2$--$\nu_3$ mixing,
which is such that $ |m_3| $ is unconstrained
by the existing results on $\nu_{\mu}$--$\nu_{\tau}$ oscillations.
The situation will change when the future results of CHORUS,
NOMAD and P803 \cite{future} become available,
which results will be able
to exclude or confirm ranges of $ |m_3| $
interesting from the cosmological point of view,
like $ |m_3| \sim 5 $ eV (hot dark matter)
or $ |m_3| \sim 50 $ eV (closure of the universe).
Meanwhile,
$ |m_3| $ can be indirectly constrained:
because $ \sin^2 (2 \theta_{e \mu}) \approx 0.04 $,
the E776 results \cite{goesgen} imply $ m_{\nu_{\mu}} < 0.6 $ eV,
and as we know the ratio $ |m_3 / m_2| $ to be around 18,
we obtain $ m_{\nu_{\tau}} < 11 $ eV.
This already excludes a substantial contribution
of the light neutrinos to the closure of the universe.

Solution 3 (see Eq.~\ref{eq:fit3})
easily accomodates the large-mixing-angle MSW solution.
In spite of the very large first-second and second-third generation
mixings,
a three-generation treatment of the MSW effect \cite{pantaleone}
does not seem essential,
because $ |K_{\nu_1 \tau}| $ remains small.
It is also possible that $ m_{\nu_{\tau}} \approx 10^{-2} $ eV
leads to MSW suppression of the GALLEX and KAMIOKANDE neutrinos,
while $ m_{\nu_{\mu}} \approx 10^{-3} $ eV
suppresses the Homestake neutrinos,
with maximal $\nu_e$--$\nu_{\mu}$ mixing \cite{wolf}.
The large second-third generation mixing implies that
$ |m_3| $ must be smaller than 0.7 eV,
with prospects of tightening this bound to 0.25 eV
in the P860 experiment \cite{future}.
This reasoning is invalidated if $ |m_3| $
is sufficiently close to $ |m_2| $:
in some solutions of type 3 those two neutrino masses
happen to be exactly equal.
But then the large mixing of $ \nu_e $ with $ \nu_2 $ and $ \nu_3 $
would imply anyway,
via the Goesgen results on missing $ \overline{\nu_e} $ \cite{goesgen},
that $ |m_3| < 0.2 $ eV.

Solutions of type 1 (see Eq.~\ref{eq:fit1})
can sometimes accomodate the large-mixing-angle MSW solution,
in which case one obtains $ |m_3| \approx 0.13 $ eV,
$ |m_2| \approx 0.005 $ eV and $ |m_1| \approx 0.003 $ eV.
An alternative possibility is the small-mixing-angle MSW effect
between $ \nu_e $ and $ \nu_{\tau} $.
Then,
$ m_3^2 \approx 4 \times 10^{-6}\, {\rm eV}^2$ and
$ m_2^2 \approx 6 \times 10^{-9}\, {\rm eV}^2 $,
and the $\nu_e$--$\nu_{\mu}$ MSW effect induces
an important additional suppression
of the neutrinos observed in the gallium experiments \cite{wolf}.
In any case,
the neutrino masses will be cosmologically very small,
and untestable by means of terrestrial oscillation experiments.

In conclusion,
the BM model is not as predictive as its authors have suggested,
because various fits to the known data are possible.
Various different possibilities for the neutrino masses
and for the lepton mixing arise.
Queer possibilities,
such as very small neutrino mass ratios or very large lepton mixings,
may occur.

\vspace{2mm}

I acknowledge discussions with Professor L.\ Wolfenstein,
who read the manuscript
and made many useful suggestions on the analysis of the neutrino spectra.
I thank Professor R.\ N.\ Mohapatra
for a short e-mail discussion on this work.
This work was supported by the United States Department of Energy,
under the contract DE-FG02-91ER-40682.

\vspace{5mm}

%
%

\begin{thebibliography}{99}
%
\bibitem{bm}
K.\ S.\ Babu and R.\ N.\ Mohapatra,
Phys.\ Rev.\ Lett.\ {\bf 70}, 2845 (1993).
%
\bibitem{seesaw}
M.\ Gell-Mann, P.\ Ramond and R.\ Slansky,
in {\it Supergravity},
edited by P.\ van Nieuwenhuizen and D.\ Freedman
(North-Holland, Amsterdam, 1979); \\
T.\ Yanagida,
in {\it Proceedings of the
Workshop on Unified Theories and Baryon Number in the Universe},
edited by O.\ Sawada and A.\ Sugamoto
(KEK, 1979).
%
\bibitem{pati}
J.\ C.\ Pati and A.\ Salam,
Phys.\ Rev.\ D {\bf 10}, 235 (1974).
%
\bibitem{models}
S.\ Dimopoulos, L.\ J.\ Hall and S.\ Raby,
Phys.\ Rev.\ D {\bf 47}, R3697 (1993);\\
K.\ S.\ Babu and Q.\ Shafi,
Phys.\ Lett.\ B {\bf 294}, 235 (1992);\\
K.\ S.\ Babu and Q.\ Shafi,
Bartol Research Institute report BA-93-09 (1993),
unpublished.
%
\bibitem{branco}
G.\ C.\ Branco and L.\ Lavoura,
Phys.\ Lett.\ B {\bf 208}, 123 (1988).
%
\bibitem{bd}
For the standard treatment of $B_d$--$\overline{B_d}$ mixing,
see for instance G.\ C.\ Branco and L.\ Lavoura,
Phys.\ Rev.\ D {\bf 38}, 2295 (1988).
%
\bibitem{gasser}
J.\ Gasser and H.\ Leutwyler,
Phys.\ Rep.\ {\bf 87}, 77 (1982).
%
\bibitem{msw}
L.\ Wolfenstein,
Phys.\ Rev.\ D {\bf 17}, 2369 (1978);\\
S.\ P.\ Mikheyev and A.\ Yu.\ Smirnov,
Yad.\ Fiz.\ {\bf 42}, 1441 (1985)
[Sov.\ J.\ Nucl.\ Phys.\ {\bf 42}, 913 (1985)].\\
For a recent combined fit of the experimental data,
see S.\ A.\ Bludman, N.\ Hata, D.\ C.\ Kennedy and P.\ G.\ Langacker,
Phys.\ Rev.\ D {\bf 47}, 2220 (1993).
%
\bibitem{mutau}
N.\ Ushida {\it et al.},
Phys.\ Rev.\ Lett.\ {\bf 57}, 2897 (1986).
%
\bibitem{future}
CHORUS Collaboration,
N.\ Armenise {\it et al.},
CERN report CERN-SPSLC/90-42 (1990),
unpublished;\\
NOMAD Collaboration,
P.\ Astier {\it et al.},
CERN report CERN-SPSLC/91-21 (1991),
unpublished;\\
Fermilab P803 and P860 proposals,
unpublished.
%
\bibitem{vacuum}
V.\ Barger, R.\ J.\ N.\ Phillips and K.\ Whisnant,
Phys.\ Rev.\ Lett.\ {\bf 65}, 3084 (1990).
%
\bibitem{goesgen}
See B.\ Blumenfeld {\it et al.},
talk given at the Neutrino '90 Conference,
CERN, 10--15 June 1990,
unpublished;
and references therein.
%
\bibitem{pantaleone}
T.\ K.\ Kuo and J.\ Pantaleone,
Phys.\ Rev.\ Lett.\ {\bf 57}, 1805 (1986),
and Phys.\ Rev.\ D {\bf 35}, 3432 (1987).
%
\bibitem{wolf}
The suggestion of this kind of double MSW effects
was first made by L.\ Wolfenstein,
Phys.\ Rev.\ D {\bf 45}, R4365 (1992).
%
\end{thebibliography}
\end{document}